\documentclass[showpacs,preprintnumbers,amsmath,amssymb,floatfix,prd]{revtex4}

\usepackage{graphicx}
\usepackage{dcolumn}
\usepackage{bm}

\newcommand{\beq}{\begin{equation}}
\newcommand{\eeq}{\end{equation}}
\newcommand{\bey}{\begin{eqnarray}}
\newcommand{\eey}{\end{eqnarray}}

\begin{document}

\title{Compact star model in Einstein Gauss-Bonnet Gravity within the framework of Finch Skea space time.}

\author{Iftikar Hossain Sardar}
\email{iftikar.spm@gmail.com } 
\affiliation{Department of Mathematics, Jadavpur University, Kolkata 700032, West Bengal, India}

\begin{abstract}

In this article we provide a new class of interior solutions of a five dimensional compact star in Einstein Gauss-Bonnet (EGB) gravity within the framework of Finch-skea space time. The Exterior space time is described by the EGB schwarzschild solution. To check physically validity  of our model we investigate various physical properties like causality of solutions, Energy conditions, mass radius relations, TOV equations etc. 


\end{abstract}

 \maketitle

\section{Introduction}
Compact star is a object which is very massive and has a small radius, i.e. it is a very high density object. In astrophysics Compact stars are used to refer collectively to white dwarfs, neutron stars and black holes. The formation of compact stars depend on usual endpoint of stellar evolution.Researchers are always very much interested to the study of relativistic models of compact stars like strange star and neutron star. Strange stars are composed of quark or strange matter while neutron stars are composed of neutrons.
In 1972, Ruderman[16] investigated the pressure inside the highly compact astrophysical objects like X-Ray buster, X-Ray pulsar, Her X-1, PSR-J1614-2230, LMC X-4 etc. having the core density beyond the nuclear density ($\sim 10^{15} g cm^{-3}$) which are anisotropic in nature. Herrera and Santos (1997)[17] studied local anisotropy in self gravitating systems extensively.
In 1984, Witten proposed the possibility of quark phase transition in the interior of a compact star. This possibility raised discussions of an entirely new class of compact stars known as strange stars. Her X-1, SAX J 1808.4-3658, PSR 0943+10, 4U 1728-34, 4U 1820-30 are some pulsars which are considered as a good strange stars.
Applying General theory of relativity one can not demonstrate the anomalous behaviour of gravitational phenomena such as the late time expansion of the universe. In view of these difficulties with the General theory of relativity, recently alternate or extended theories of gravity have aroused with  great interest. The mathematical reason for this interest is that the higher order derivative curvature terms make a nonzero contribution  to the dynamics. The behaviour and the dynamics of the gravitational field can be extended into higher dimensions easily. Einstein Gauss-Bonnet gravity is proved promising in this regard and therefore widely studied further.\\ \\
~~~In string theory Einstein Gauss-Bonnet gravity  appears naturally to consider effective action in low energy limit. This gravity is the generalization of Einstein gravity by adding an extra term with the standard Einstein Hilbert action, which is quadratic in the Riemann tensor. Varying this extra term with respect to the metric only one can generate a system of second order equations and thus the theory shares many nice properties of classical general relativity. In 4-D, Einstein Gauss-Bonnet gravity and general relativity  are equivalent. Dealing with the aspects of gravitational collapse, EGB gravity indicates several new results. The study of causal structure of the singularities in EGB is different from classical general relativity for spherically symmetric inhomogeneous distribution of dust and null dust. Many interesting results in EGB theory regarding black hole models have been widely studied. Boulware and Deser[10] generalized the higher dimensional solutions in Einstein theory due to Tangherlini[14] and  the contribution of the EGB theory with quadratic curvature terms have been included by Myers and Perry[15]. However Dadhich et al.[11], Jhingan and Ghosh[13] have found the explicit exact solutions which have showed that Schwarzschild interior solutions is universal in the sense and it is valid in both the EGB gravity and higher dimensional Einstein theory.
~~~In general relativity, modelling of a compact stars with static spherically symmetric metric have been studied extensively in last century. Many researchers have been found various exact solutions of Einstein and Einstein-Maxwell system of field equations for the both charged and uncharged system of matter distribution. For a physically valid model, It should satisfy the physical properties like the energy conditions, stability with respect to the radial perturbations. Also the speed of sound velocity within the fluid distribution must be less than the speed of light and the compactness of the star should satisfy the Buchdahl's inequality.
Recently Hansraj, Chilambwe and Maharaj[2] find out the exact solutions for  spherical static perfect fluid and Bhar et al. describe the comparative study between Einstein gravity and Einstein Gauss-Bonnet gravity in Krori-Barua spacetime.  Inspiring by these models I am trying to develop a model of compact star in EGB gravity within the framework of Finch-Skea spacetime. In this work I have tried to study various physical properties to check the physical validity of the model. Several authors (Rahaman et al. 2010,2012,2015; Abbas et al. 2014)[3],[8] used Herrera's cracking (or overturning) concept to check stability of the star. In this regard I have also checked whether the model satisfies the causality condition or not.  \\ \\
~~~ The plan of this paper is as follows: In Sect.2 we have discussed about the Einstein Gauss-Bonnet gravity. In Sect.3  and Sect.4  we have studied field equations and the mathematical solutions of EGB gravity within finch-Skea space time. In Sect.5 exterior space time and matching conditions have been studied. Several physical properties have been studied in Sect.6. Finally in the last Sect.7 we have provided a concluding remarks regarding our model.\\

\section{Einstein Gauss-Bonnet Gravity}
The action in five dimensional Gauss-Bonnet gravity is written as,
\begin{equation}
s=\int \sqrt{-g}[\frac{1}{2}(R-2\Lambda+\alpha L_{GB})]d^5x+\mathcal{S}_{M}
\end{equation}
The first term is the Einstein-Hilbert action and the last term is matter Lagrangian. In this action $\alpha$ represents Gauss-Bonnet coupling constant. The Gauss-Bonnet Lagrangian is quadratic in Ricci scalar, Ricci tensor and the Riemann tensor, which is given by
\begin{equation}
L_{GB}=R^2+R_{abcd}R^{abcd}-4R_{cd}R^{cd}
\end{equation}
 However the action has an advantage that the equation of motion turn out to be second order quasi linear which is consistent to the Einstein gravity. The Gauss-Bonnet term makes no influence for $n\leq4$ but for $n>4$ it has contribution.\\
  
  The EGB field equations can be written in the form,
\begin{equation}
G_{ab}+\alpha H_{ab}=T_{ab}
\end{equation}
with metric signature (- + + + +). Here the quantities $G_{ab}$,  $H_{ab}$ and ${T_{ab}}$ are the Einstein tensor, Lanczos tensor and Energy momentum tensor respectively. The Lanczos tensor is defined in the form,
\begin{equation}
H_{ab}=2(RR_{ab}-2R_{ac}R^c_b-2R^{cd}R_{acbd}+R^{cde}_aR_{bcde})-\frac{1}{2}g_{ab}L_{GB}
\end{equation}

According to low energy effective action in string theory, the coupling constant $\alpha$ is related to the inverse string tension which is positive definite. The condition $\alpha\geq0$ is generally considered. But in this paper we consider the condition for $\alpha>0$ only. We also consider geometric units k=1.

\section{Field Equations}
The metric for static spherically symmetric in five dimensional space time is taken as,
\begin{equation}
ds^2 = -e^{2\nu(r)}dt^2+e^{2\lambda(r)}dr^2+r^2(d\theta^2+sin^2\theta d\phi^2+sin^2\theta sin^2\phi d\psi^2)
\end{equation}
in coordinates $(x^i=t,r,\theta,\phi,\psi)$, where $\nu(r)$ and $\lambda(r)$ are the gravitational potentials. Here we consider the matter field is a perfect fluid with energy momentum tensor $T_{ab}=(\rho+p)u_au_b+pg_{ab}$ and the comoving fluid velocity as $u^a=e^{-\nu}\delta^a_0$.
By the EGB field equation (3) reduces to
\begin{eqnarray}
\rho & = & \frac{3}{e^{4\lambda}r^3}(4\alpha \lambda^{\prime}+r e^{2\lambda}-r e^{4\lambda}-r^2 e^{2\lambda}\lambda^{\prime}-4\alpha e^{2\lambda}\lambda^{\prime})
\\
p_r & = & \frac{3}{e^{4\lambda}r^3}[-r e^{4\lambda}+(r^2\nu^{\prime}+r+4\alpha \nu^{\prime})e^{2\lambda}-4\alpha \nu^\prime]
\\
p_t & = & \frac{1}{e^{4\lambda}r^2}(-e^{4\lambda}-4\alpha \nu^{\prime\prime}+12\alpha\lambda^\prime \nu^\prime-4\alpha(\nu^\prime)^2)\nonumber\\
& + & \frac{1}{e^{2\lambda}r^2}(1-r^2\lambda^\prime \nu^\prime+2r\nu^\prime-2r\lambda^\prime+r^2(\nu^\prime)^2)\nonumber\\
& + & \frac{1}{e^{2\lambda}r^2}(r^2\nu^{\prime\prime}-4\alpha\lambda^\prime\nu^\prime+4\alpha(\nu^\prime)^2+4\alpha\nu^{\prime\prime});
\end{eqnarray}
Where $ \rho $, $ p_r $, $ p_t $ denotes the matter density, radial pressure, transverse pressure respectively and $ ^\prime $ denotes the differentiation w.r.to radial coordinate r.



\section{Solutions}
To explore nature of the physical parameters, we consider the metric potential in Finch Skea space time solutions is given by
\begin{equation}
e^{2\lambda(r)} = \left( 1+\frac{r^2}{R^2}\right) 
\end{equation}
where R is the constant.\\
Solving equations (6) to (8) using (9), we get
\begin{eqnarray}
\rho & = & \frac{3R^2}{(R^2+r^2)^2}\left[-2-\frac{r^2}{R^2}-\frac{4\alpha}{(R^2+r^2)}\right]
\\
p_r & = & \frac{3R^2}{r^3(R^2+r^2)}\left[ -\frac{r^3}{R^2}+\nu^\prime\left( r^2+4\alpha-\frac{4\alpha R^2}{R^2+r^2}\right) \right]
\\
p_t & = & \frac{R^4}{r^2(R^2+r^2)^2}\left[ -\left( \frac{R^2+r^2}{R^2}\right) ^2-4\alpha\nu^{\prime\prime}+\nu^\prime\frac{12\alpha r}{R^2+r^2}-4\alpha(\nu^\prime)^2\right] \nonumber\\
& + & \frac{R^2}{r^2(R^2+r^2)}\left[ 1-\nu^\prime \frac{r^3}{R^2+r^2}+2r\nu^\prime-\frac{2r^2}{R^2+r^2}+r^2(\nu^\prime)^2\right] \nonumber\\
& + & \frac{R^2}{r^2(R^2+r^2)}\left[r^2\nu^{\prime\prime}-\frac{4\alpha r}{R^2+r^2}\nu^\prime+4\alpha(\nu^\prime)^2+4\alpha\nu^{\prime\prime}\right] 
\end{eqnarray}
Further we consider the EOS having the form\\
\begin{equation}
p_r = \omega\rho
\end{equation}
Where $\omega$ is the equation of state parameter. Now equations (10) and (11) lead to
\begin{equation}
\nu^\prime  =  \frac{r}{(4\alpha+R^2+r^2)}\left[ 1+\frac{r^2}{R^2}-\omega(2+\frac{r^2}{R^2}+\frac{4\alpha}{R^2+r^2})\right]
\end{equation}
Solving this equation we get
\begin{equation}
\nu = \left(\frac{r^2 - 4\alpha ln(4\alpha +R^2+r^2)}{2R^2}\right)(1-\omega)-\frac{\omega ln(R^2+r^2)}{2}+\nu_0
\end{equation}
Where $\nu_0$ is an integrating constant Which can be choosen equal to 0, without any loss of generality.\\

 Therefore the other parameters are obtained in the following form \\
\begin{eqnarray}
\rho & = & \frac{3R^2}{(R^2+r^2)^2}\left[-2-\frac{r^2}{R^2}-\frac{4\alpha}{(R^2+r^2)}\right]
\\
p_r & = & \frac{3R^2 \omega}{(R^2+r^2)^2}\left[-2-\frac{r^2}{R^2}-\frac{4\alpha}{(R^2+r^2)}\right]
\\
p_t & = & -\frac{1}{R^2+r^2}+\left( \frac{R^2}{R^2+r^2}+\frac{4\alpha R^2}{(R^2+r^2)^2}\right)(\nu^{\prime\prime}+(\nu^\prime)^2)\nonumber\\
& + & \frac{r}{R^2+r^2}\nu^\prime \left[ \frac{4\alpha R^2(2R^2-r^2)}{r^2(R^2+r^2)^2}-\frac{R^2}{R^2+r^2}\right]
+\frac{2R^2}{r(R^2+r^2)}\left( \nu^\prime-\frac{r}{R^2+r^2}\right) 
\end{eqnarray}
Where 
\begin{eqnarray}
\nu^\prime & = & \frac{r}{(4\alpha+R^2+r^2)}\left[ 1+\frac{r^2}{R^2}-\omega(2+\frac{r^2}{R^2}+\frac{4\alpha}{R^2+r^2})\right]\nonumber\\ 
\nu^{\prime\prime} & = & \frac{\nu^\prime}{r}-\frac{2r\nu^\prime}{4\alpha+R^2+r^2}+\frac{r\left[\frac{2r}{R^2}-\omega\left(\frac{2r}{R^2}-\frac{8\alpha r}{(R^2+r^2)^2} \right)  \right] }{4\alpha+R^2+r^2}\nonumber
\end{eqnarray}
We have shown the characteristics of the parameters $\rho$, $p_r$ and $p_t$ in fig. 1. In the fig.(1),  left panel indicates  that matter density is a decreasing function and in the right panel both the radial pressure and tangential pressure are decreasing functions.In fig.(2), We also note that the anisotropy of the model is vanishes at origin i.e, both radial pressure and tangential pressure are equal at origin which indicates that the solutions are regular at origin.
\begin{figure}[thbp]
\includegraphics[scale=.3]{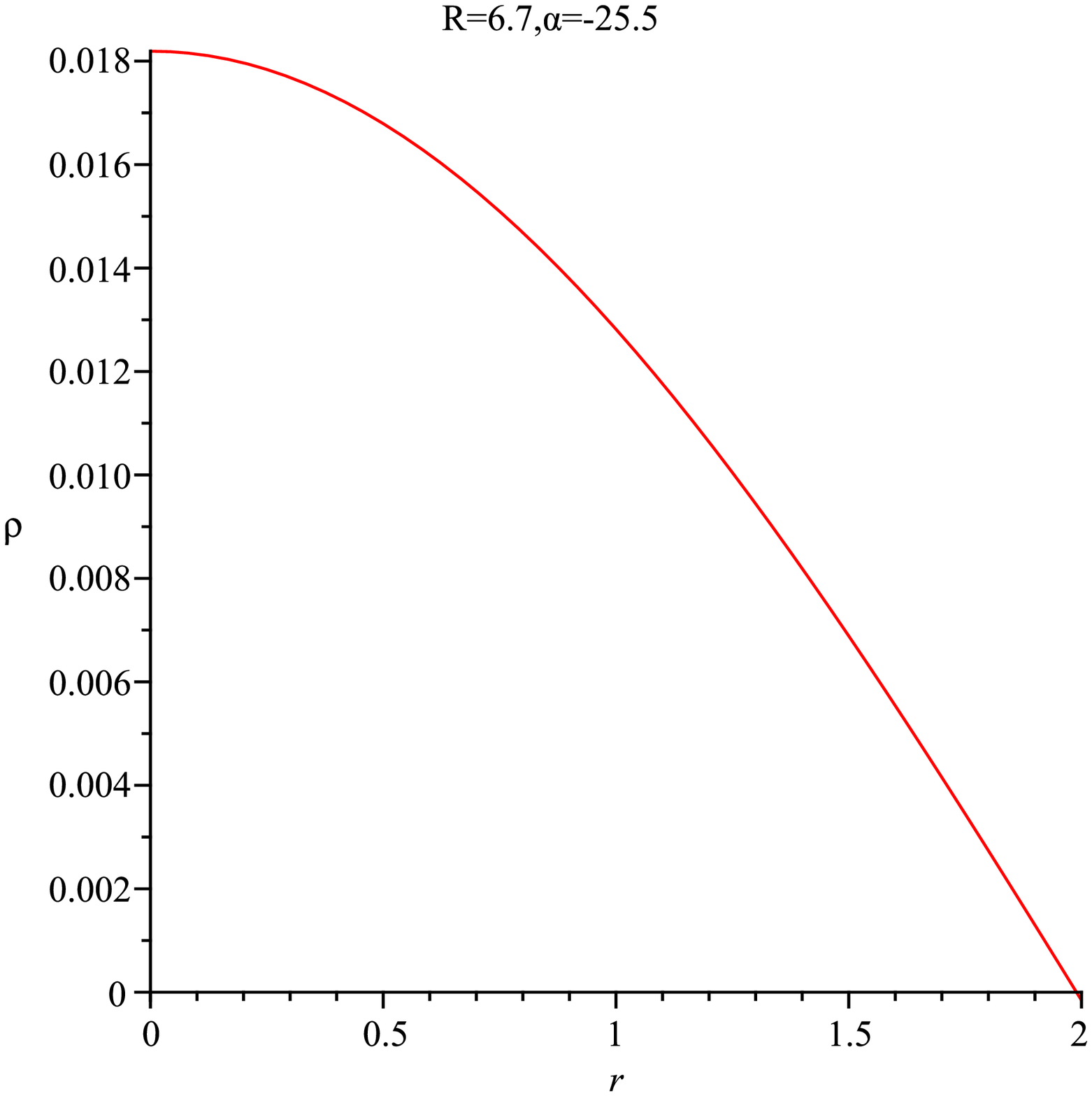}
\includegraphics[scale=.3]{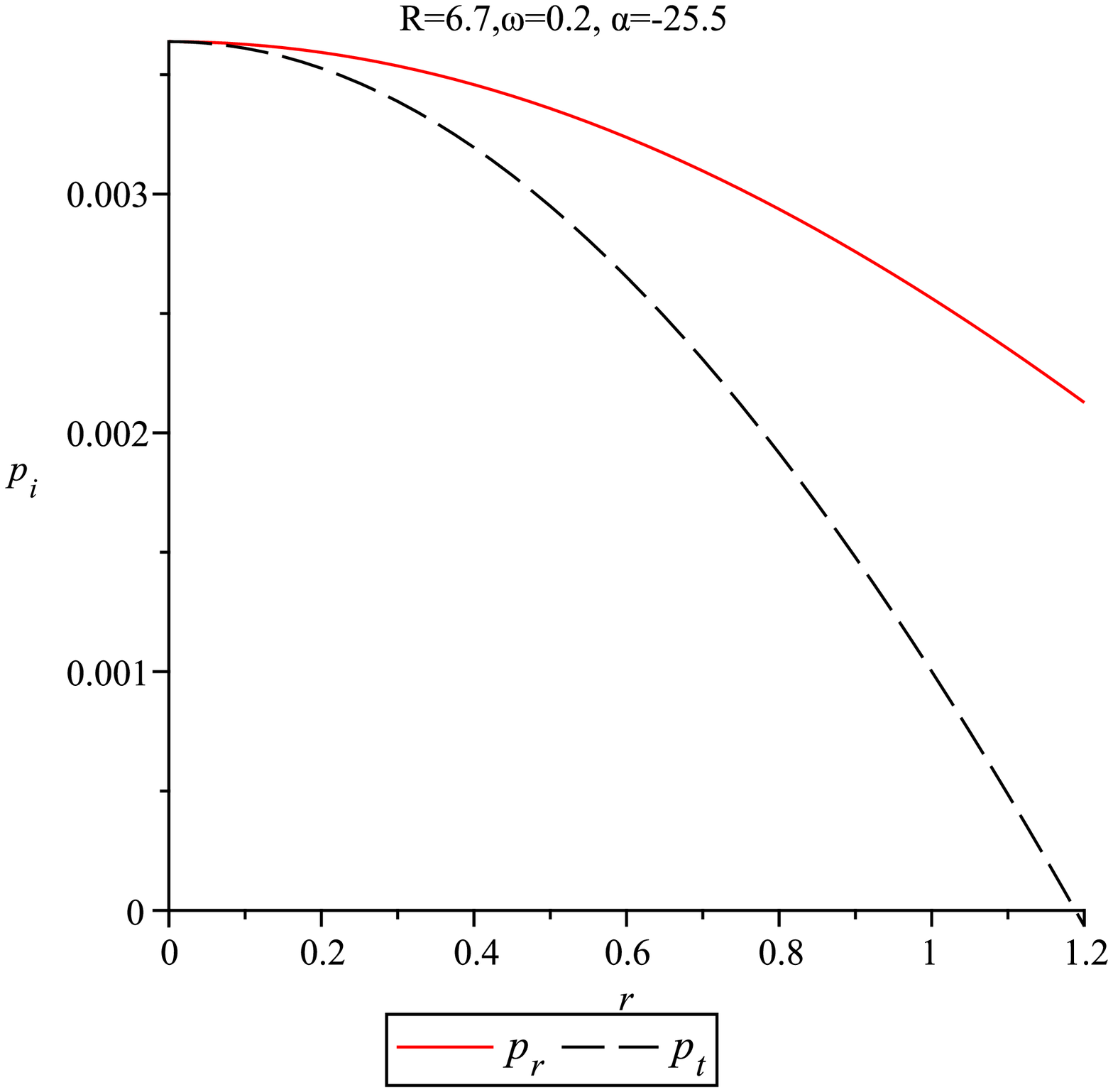}
\caption{(Left) Variation of Matter density is plotted against radial coordinate r.  (Right)  Variation of radial and tangential pressures  are plotted against radial coordinate r}
\end{figure}

\begin{figure}[thbp]
\centering
\includegraphics[scale=0.4]{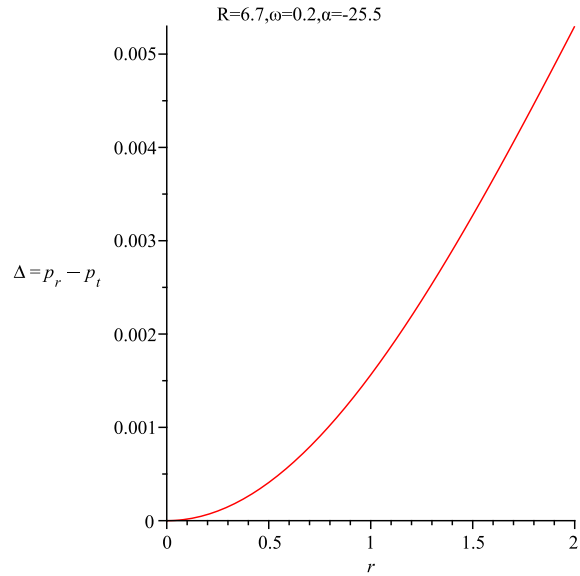}
\caption{ Variation of Anisotropic factor $\Delta$ is plotted against radial coordinate r. }
\end{figure}
\section{Exterior space time and matching conditions}
In five dimensional stellar model Einstein Gauss Bonnet(EGB) Schwarzschild solution [10] described the static exterior space time, which is given by
\begin{equation}
ds^2 = -F(r)dt^2 + [F(r)]^{-1} dr^2 + r^2(d{\theta}^2 +sin^2{\theta} d{\phi}^2 + sin^2{\theta}sin^2{\phi} d{\psi}^2 )
\end{equation}
Where $F(r)$ is given by 
\begin{equation}
F(r) = 1+\frac{r^2}{4\alpha}\left(1-\sqrt{1+\frac{8\alpha M}{r^4}}\right)
\end{equation}
where M is connected with the gravitational mass of the stellar model.\\
Using the continuity of the metric functions and their derivatives over the boundary $r = a(Radius)$ we get\\
\begin{center}
$g_{tt}(Exterior) = g_{tt}(Interior)$
\end{center} 
and 
\begin{center}
 $g_{rr}(Exterior) = g_{rr}(Interior)$
 \end{center} 
i.e,\\
\begin{eqnarray}
e^{\left[\left(\frac{a^2-4\alpha ln(4\alpha+R^2+a^2)}{R^2}\right)(1-\omega)-\omega ln(R^2+a^2)+\nu_0\right]} & = & 1+\frac{a^2}{4\alpha}\left(1-\sqrt{1+\frac{8\alpha M}{a^4}}\right)\nonumber\\
1+\frac{a^2}{R^2} & = & \left[1+\frac{a^2}{4\alpha}\left(1-\sqrt{1+\frac{8\alpha M}{a^4}}\right)\right]^{-1}\nonumber
\end{eqnarray}
From the above set of equations we get the solutions as
\begin{eqnarray}
R^2 &=& \frac{4\alpha}{\sqrt{1+\frac{8\alpha M}{a^4}}-1} - a^2
\\
\nu_0 &=& \omega ln(R^2 + a^2)+\frac{a^2 -4\alpha ln(4\alpha +R^2 + a^2)}{R^2}(\omega - 1) + ln \left[1+\frac{a^2}{4\alpha}\left(1-\sqrt{1+\frac{8\alpha M}{a^4}}\right)\right]
\end{eqnarray}
\section{Physical properties}
\subsection{Energy conditions}
It is well known that the anisotropic stellar model will be satisfied the null energy condition(NEC), weak energy condition(WEC) and strong energy condition(SEC) if and only if the following equations hold simultaneously:
\begin{eqnarray}
\rho\geq 0
\\
\rho + p_r \geq 0,  \rho + p_t \geq 0
\\
\rho + p_r +2p_t\geq 0
\end{eqnarray}
In fig.(3) we have shown that all the energy conditions are satisfied by our model.

\begin{figure}[thbp]
\centering
\includegraphics[scale=0.4]{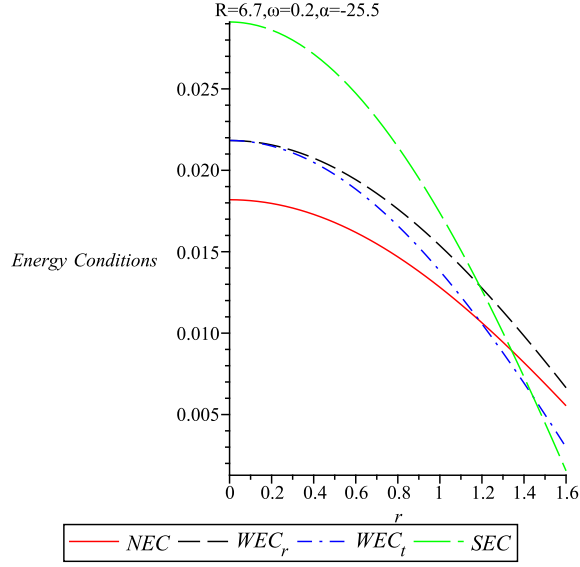}
\caption{ The Variation of the Energy conditions of the stellar object is plotted against radial coordinate r.}
\end{figure}
\subsection{Mass radius relation }
The effective mass of the stellar object within the radial distance r is given by,
\begin{equation}
M_{eff} = {\int}_{0}^{r} 2{\pi}^2 r^3 \rho dr
        = 6{\pi}^2 R^2 {\int}_{0}^{r} \frac{r^3}{(R^2+r^2)^2} (-2-\frac{r^2}{R^2}-\frac{4\alpha}{(R^2+r^2)^2})
\end{equation}
Now, the compactness of the stellar object is given by the relation 
\begin{equation}
u = \frac{M_{eff}}{r} = 6{\pi}^2 R^2 {\int}_{0}^{r} \frac{r^2}{(R^2+r^2)^2} (-2-\frac{r^2}{R^2}-\frac{4\alpha}{(R^2+r^2)^2})
\end{equation}
The variation of mass and compactness of the stellar object are illustrated graphically. In fig.(4) we have seen that compactness(u) is an increasing function of r.

\begin{figure}[thbp]
\centering
\includegraphics[scale=.3]{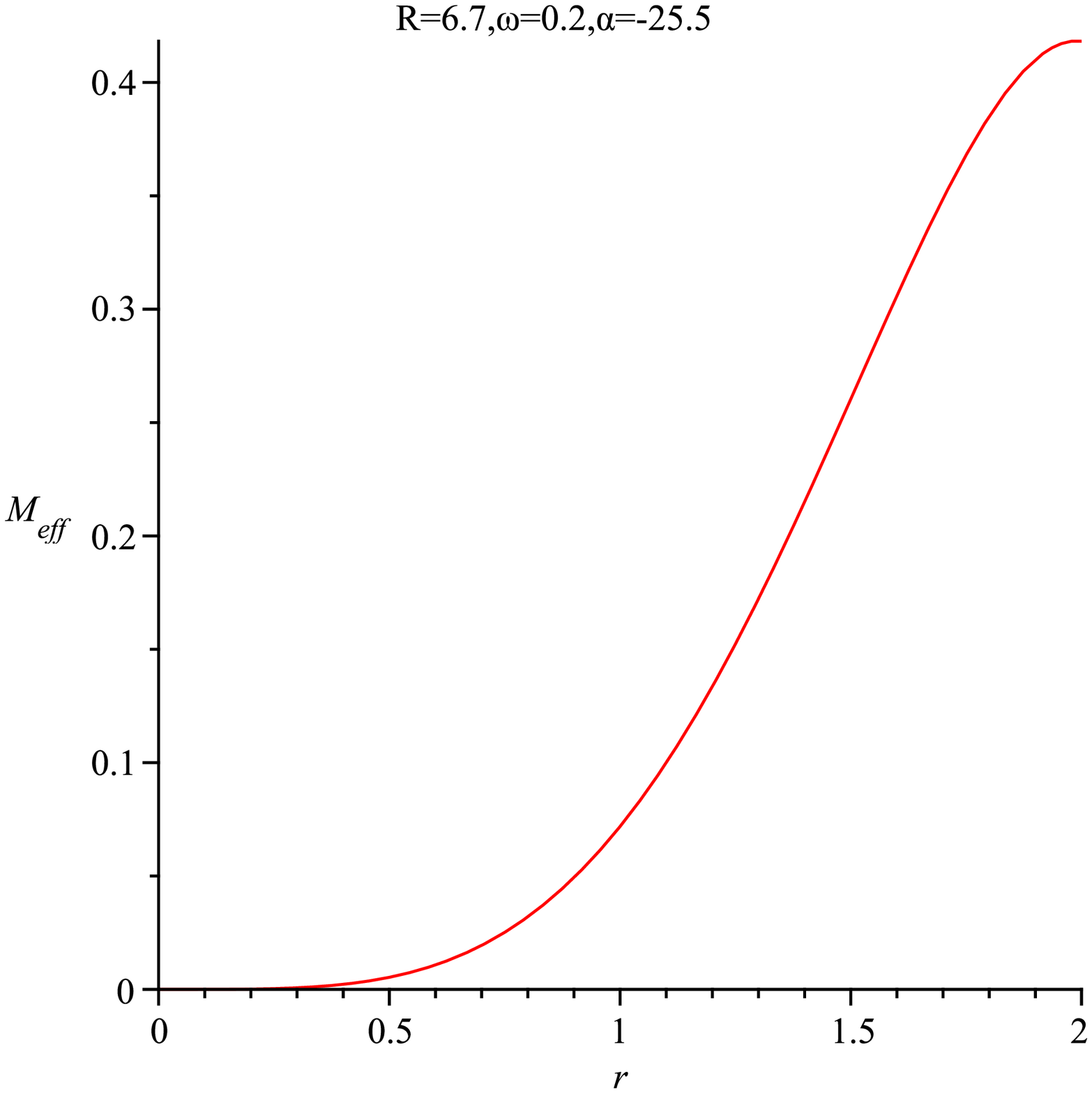}
\includegraphics[scale=.3]{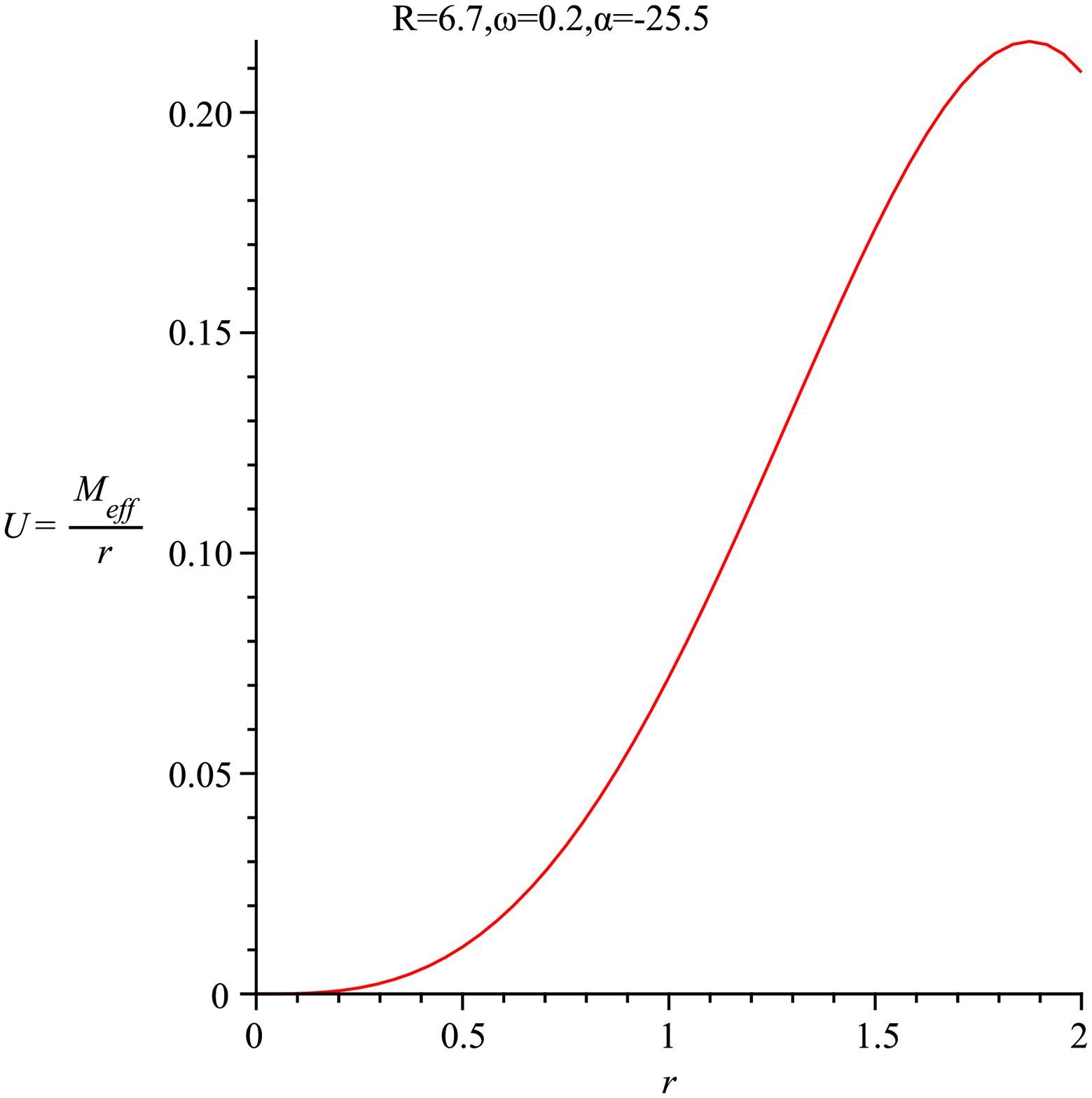}
\caption{(Left)  Variation of Mass of the stellar object is  plotted against radial coordinate r.
 (Right)   Compactness of the stellar object  is  plotted against radial coordinate r.}
\end{figure}
\subsection{Surface Red shift}
The surface red shift function $Z_s$ of the stellar object is given by 
\begin{equation}
Z_s = (1 - 2u)^{-\frac{1}{2}} - 1
    = \left(1 - 12{\pi}^2 R^2 {\int}_{0}^{r} \frac{r^2}{(R^2+r^2)^2} (-2-\frac{r^2}{R^2}-\frac{4\alpha}{(R^2+r^2)^2})\right)^{-\frac{1}{2}} - 1
\end{equation}
which has been illustrated graphically in fig.(5). 

\begin{figure}[thbp]
\centering
\includegraphics[scale=0.4]{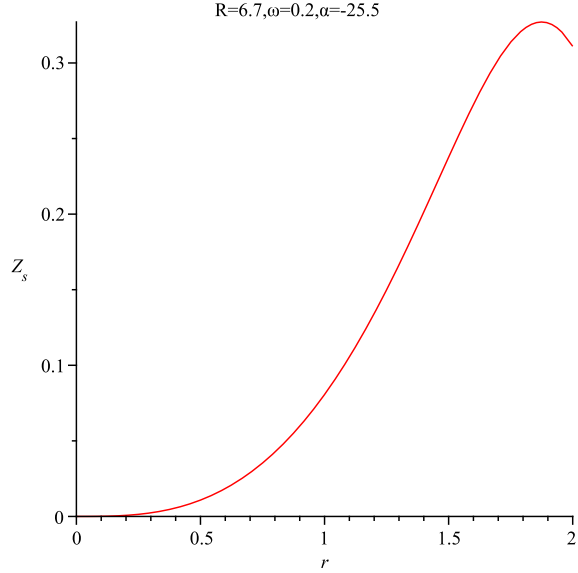}
\caption{ Surface redshift function of the stellar object is plotted against radial coordinate r.}
\end{figure}

\subsection{TOV Equations}
The generalized Tolman-Oppenheimer-Volkoff(TOV) equations can be written in the following form,
\begin{equation}
-\frac{M_G(\rho + p_r)}{r^2} e^{\frac{\lambda - \nu}{2}} - \frac{dp_r}{dr} + \frac{2}{r} (p_t - p_r) = 0
\end{equation}
Where $M_G = M_G (r)$ is the effective gravitational mass inside the sphere with radius r. The effective gravitational mass is given by the following expression
\begin{equation}
M_G (r) = \frac{1}{2}r^2 e^{\frac{\lambda - \nu}{2}} {\nu}^\prime
\end{equation}
The TOV equation describes the equilibrium of the stellar configuration subject to the forces such as gravitational$(F_g)$, hydrostatic$(F_h)$ and anisotropic stress$(F_a)$ so that
\begin{equation}
F_g + F_h + F_a = 0
\end{equation}
Where 
\begin{eqnarray}
F_g & = & -\frac{{\nu}^\prime}{2}(\rho + p_r) \\
F_h & = & -\frac{dp_r}{dr}\\
F_a & = & \frac{2}{r} (p_t - p_r)
\end{eqnarray}
These forces are graphically plotted in fig.(6), which describes the equilibrium of the stellar configuration under the combined effects of these forces.

\begin{figure}[thbp]
\centering
\includegraphics[scale=0.4]{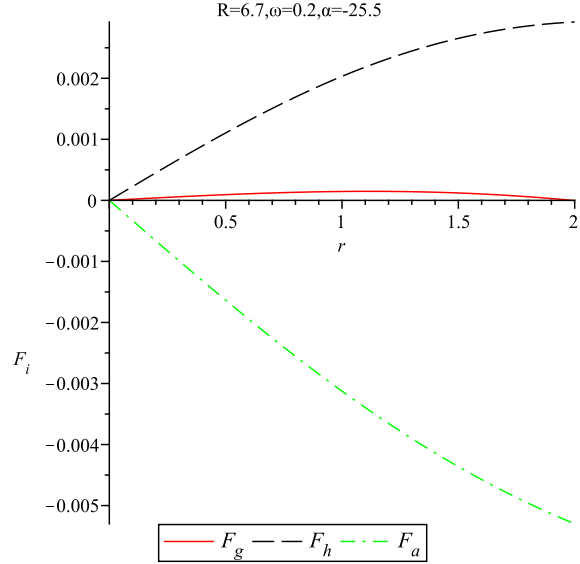}
\caption{ Contribution of different forces acting on interior of the stellar object in static equilibrium.}
\end{figure}


\begin{figure}[thbp]
\centering
\includegraphics[scale=.3]{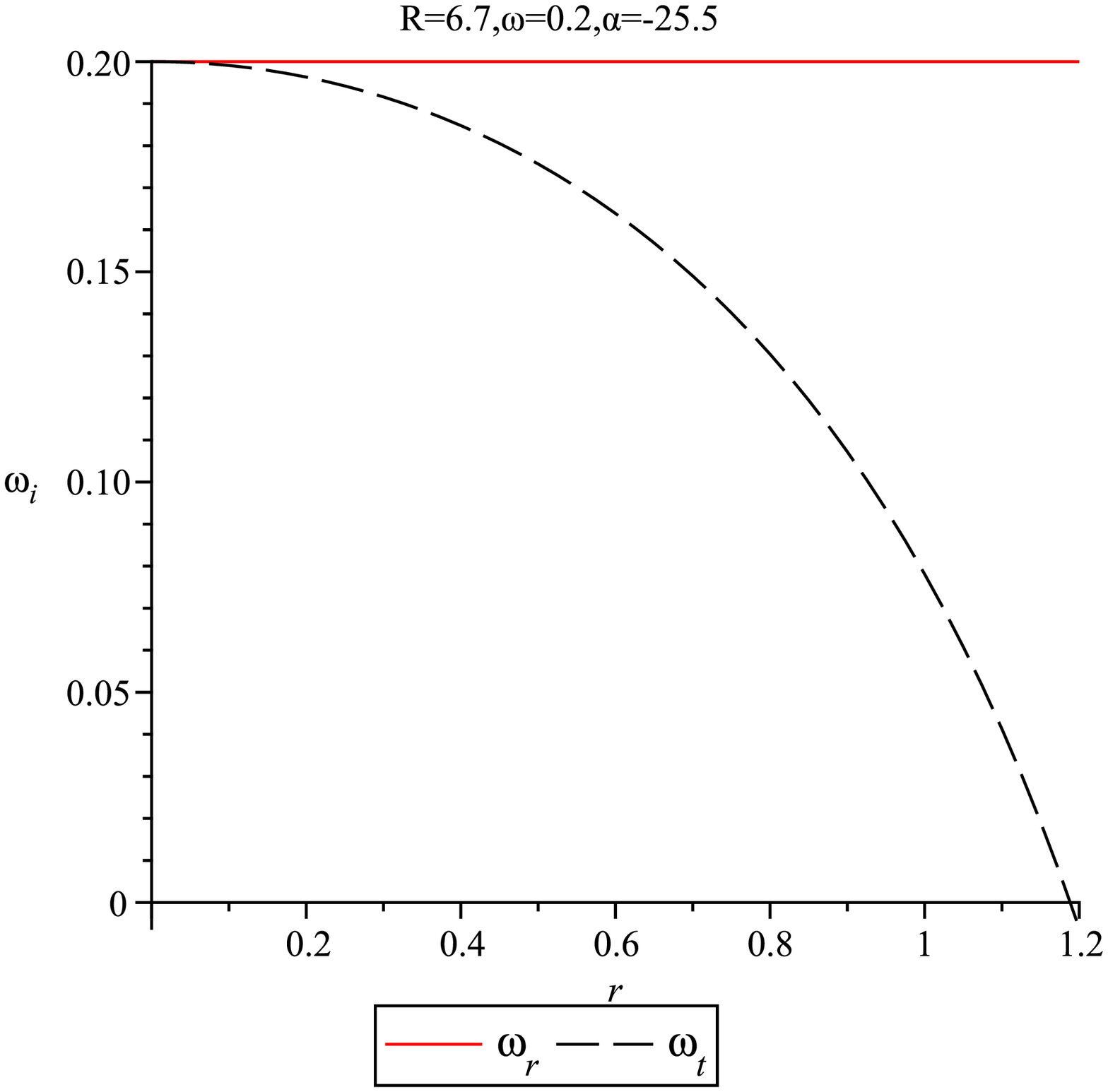}
\includegraphics[scale=.3]{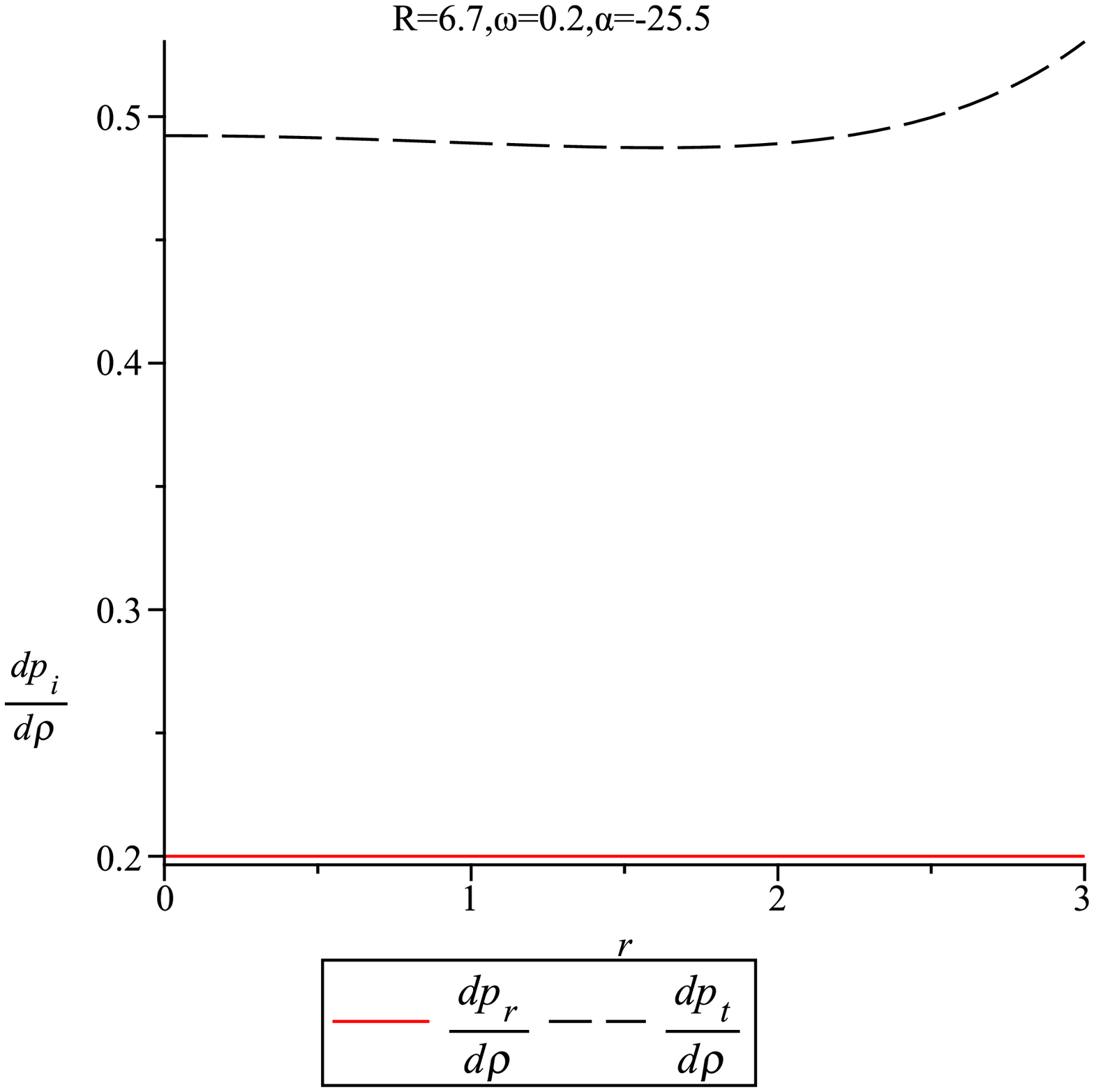}
\caption{(Left)    Variation of Equation state parameter is  plotted against radial coordinate r.
 (Right)  Sound velocity of the stellar object  is  plotted against radial coordinate r.}
\end{figure}

\pagebreak
\section{Concluding Remarks}
In the present work we have proposed a new class of interior solutions of compact object in five dimensional Einstein Gauss Bonnet(EGB) gravity with the help of Finch-Skea metric solution. The solutions are regular at the centre and satisfy all the physical requirements. The mass function satisfy the buchdahl's inequality (maximally allowable mass radius ratio $\frac{2M}{r} < \frac{8}{9}$)[6]. In the model the maximally allowable mass radius ratio is $\simeq 0.21022 < \frac{4}{9}$,  which is within the standard limit. Radial and tangential EoS parameter have been plotted In fig.(7), which shows that both the radial and the tangential EoS  parameter lies between 0 and 1. i.e, $0 < {\omega}_r <1$ and $0 < {\omega}_t <1$, therefore from [4,5] we have $\mid v_{st}^2 - v_{sr}^2 \mid \leq 1$.\\
i.e;
\begin{center}
$-1\leq v_{st}^2 - v_{sr}^2 \leq 1$
\end{center}
which shows that the stellar object satisfies the causal condition. Plugging G and c in appropriate places one can determined the central density and pressure. In the model the central density and the central pressure are obtained as $24.6435739965\times 10^{15} gm cm^{-3}$ and $91.405263273\times 10^{35} dyne cm^{-2}$ respectively. We also calculate the surface redshift as $Z_s = 0.31355$, which shows that the model is highly compact object. From equation (26), we calculate the effective mass of the compact model as $0.2836146311 M_0$. For all these calculations we consider the radius of the star is 1.99 km and $R = 6.7$, $\omega = 0.2$ and $\alpha = -25.5$.

\section{Acknowledgements}
I am thankful to DST, The Government of India for providing financial support under the INSPIRE Fellowship.

\section{References}
\bibliographystyle{plain} 
 $[1]$ Bhar.P, Govender.M, Sharma.R , Arxiv:1607.06664v1 [gr-qc] 21 Jul (2016).\\
$[2]$ Hansraj.S, Chilambwe.B, Maharaj.S.D, Eur.Phys. J.C {\bf 75}, (2015) 277.\\
$[3]$ Varela.V, Rahaman.F, Ray.S, Chakraborty.K and Kalam.M, Phys. Rev. D {\bf 82}, (2010) 044052.\\
$[4]$ Herrera.L, Phys. Lett. A {\bf 165}, 206 (1992).\\ 
$[5]$ Andreasson.H, Commun. Math. Phys, {\bf 288}, 715 (2009)\\
$[6]$ Islam.S, Rahaman.F, Sardar.I.H, Astrophys. Space. Sci,{\bf 356}, (2015) 293-300.\\
$[7]$ Buchdahl.H.A, Phys. Rev. {\bf 116}, 1027 (1959).\\
$[8]$ Finch.M.R, Skea.J.E.F, Class. Quanrum. Gravity {\bf 6}, 467 (1989).\\
$[9]$ Abbas.G et. al: Astrophys. Space. Sci, {\bf 354}, 2110 (2014).\\
$[10]$ Bhar.P, Rahaman.F, Biswas.R, Fatima.H.I, Commun. Theor. Phys. {\bf 62}, (2014) 221-226. \\
$[11]$ Boulware.D.G, Deser.S, Phys.Rev.Lett. {\bf 55}, 2656,(1985).\\
$[12]$ Dadhich.N.K, Molina.A, Khugaev.A,: Phys. Rev. D {\bf 81}, 104026 (2010). \\
$[13]$ Tolman.R.C, Phys.Rev. {\bf 55}, 364 (1939).\\
$[14]$ Jhingan.S, Ghosh.S.G,: Phys. Rev. D {\bf 81}, 024010 (2010). \\
$[15]$ Tangherlini.F.R,: I1 Nuovo Cimento, {\bf 27}, 636 (1963). \\
$[16]$ Myers.R.C, Perry.M.J,: Ann.Phys.{\bf 172}, 304 (1986).\\
$[17]$ Ruderman.R,: Annu. Rev. Astron. Astrophys. {\bf 10}, 427 (1972).\\
$[18]$ Herrera.L, Santos.N.O,: Phys. Rep. {\bf 286}, 53 (1997).\\


\end{document}